\documentclass[structabstract]{aa}  
\usepackage{graphicx}
\usepackage{natbib}
\usepackage{txfonts}
\begin{document}
   \title{The new Wolf-Rayet binary system WR62a}

   %\subtitle{I. Overviewing the $\kappa$-mechanism}

   \author{A. Collado
          \inst{1}
\fnmsep\thanks{Fellow of CONICET, and visiting astronomer, Cerro Tololo
    Inter-American Observatory (Chile) and CASLEO (Argentina).
}, 
          %\and
          R. Gamen\inst{2}\fnmsep\thanks{Visiting astronomer, Cerro Tololo
    Inter-American Observatory (Chile) and CASLEO (Argentina).
}
          \and
          R.H. Barb\'a\inst{1,3}\fnmsep\thanks{Visiting astronomer, Las Campanas
            Observatory and Cerro Tololo Inter-American Observatory, Chile.}
          }

\institute{
Instituto de Ciencias Astron\'omicas de la Tierra y del Espacio (ICATE), 
CONICET, Avda. Espa\~na 1512 Sur, J5402DSP, San Juan, Argentina
%            \and Facultad de Ciencias Exactas F\'isicas y Naturales (UNSJ), San Juan, Argentina
            \and 
Instituto de Astrof\'isica de La Plata, CONICET; Facultad de Ciencias
Astron\'omicas y Geof\'{\i}sicas, Universidad Nacional de La Plata, 
Paseo del bosque s/n, B1900FWA, La Plata, Argentina
            \and
Departamento de F\'{\i}sica, Universidad de La Serena, Av. Cisternas 1200
Norte, La Serena, Chile\\  
      \email{acollado@icate-conicet.gob.ar}
             %\thanks{The university of heaven temporarily does not
              %       accept e-mails}
             }

   \date{Received; accepted}

% \abstract{}{}{}{}{} 
% 5 {} token are mandatory
 
\abstract
% context heading (optional)
 {A significant number of the Wolf-Rayet stars seem to be binary or
   multiple systems, but the nature of many of them is still unknown. 
 Dedicated monitoring of WR stars favours the discovery of new systems.} 
% aims heading (mandatory)
 {We explore the possibility that WR62a is a binary system.}
 {We analysed the spectra of WR62a, obtained between 2002 and 2010, 
  to look for radial-velocity and spectral variations that would suggest there
  is a binary component.
  We searched for periodicities in the measured radial velocities and 
  determined orbital solutions. A 
  period search was also performed on the ``All-Sky Automated Survey''
  photometry.} 
% results heading (mandatory)
  {We find that WR62a is a double-lined spectroscopic binary with a
        WN5 primary star and an O\,5.5-6 type secondary component 
        in orbit with a period of 9.1447 d. 
The minimum masses range between 21 and 23 $M_\odot$ for the WN star and 
between 39 and 42 $M_\odot$ for the O-type star,  
thus indicating that the WN star is less massive than the O-type
component.  
 We detect a phase shift in the radial-velocity curve of the 
 He\,{\sc ii} $\lambda$4686 emission line relative to the other emission
 line curves.  
The equivalent width of this emission line shows a minimum value when the WN
star passes in front of the system.  
The analysis of the ASAS photometry confirms the spectroscopic periodicity, 
presenting a minimum at the same phase.}
  % conclusions heading (optional), leave it empty if necessary 
{}

\keywords{Stars: binaries: spectroscopic --
                Stars: individual: WR62a --
                Stars: Wolf-Rayet
               }
\authorrunning{Collado et al.}
\titlerunning{The new Wolf-Rayet binary system WR62a}

   \maketitle
%
%________________________________________________________________

\section{Introduction}

Wolf-Rayet stars frequently occur in binary systems. In fact, multiplicity
seems to be one of the most striking properties of massive stars
\citep[cf. ][]{2011IAUS..272..474S}.   
\citet{huc01} has found 38\% binaries and possible binaries among a total of
227 WR stars. 
\citet{Langer_1999} argue that most of the Supernova Type Ib/c occur in WR + OB 
interacting binary systems, and given the statistics derived from the SN
observed, it would be expected that more WR stars are detected in binary
systems than as single stars.  
Another reason is provided by the observation of late-type WC stars (WC8~-~9). 
These stars exhibit thermal emission from hot dust, and it is generally
accepted that the production of dust is a consequence of the binarity,
  where the dust is created via the collision of hydrogen-rich and
carbon-rich winds from the respective OB and WC stars
\citep{1991MNRAS.252...49U}.  
Moreover, a systematic high-resolution spectroscopic survey of O and WN stars
\citep[{\em OWN Survey}; ][]{2010RMxAC..38...30B} has found that nearly 60\%
of the sample are candidates to belong in binary or multiple systems. 
Therefore, the binary fraction is expected to be higher than what was found
by van der Hucht, that is, the binarity appears to play a key role in the
formation of WR stars. In recent years, the discovery of new binary systems
has been favoured thanks to technological advances, allowing us to understand
a little more about the nature of massive stars. 

The stellar mass is a fundamental astrophysical parameter whose knowledge,
together with the rate of mass loss, chemical composition, and rotation,
determines the properties and evolution of stars \citep{Meynet_2005}.  
Observing binaries, particularly double-lined spectroscopic binary systems
(SB2) that are eclipsing, is the most direct way to obtain accurate stellar
masses. 
 
We are carrying out spectroscopic monitoring of faint southern galactic
Wolf-Rayet stars in order to discover new binary systems. One of the most
conspicuous radial-velocity (RV) variables observed in our sample is
WR62a. This star (SMSNPL11, $\alpha_{2000}=14^h32^m38\fs18$;
$\delta_{2000}=-61^{\circ}29'55\farcs4$; $v$=13.8) was discovered by
\citet{Shara_1999}. They classified it as WN4-5o, which means that they did
not identify absorption lines in the spectrum.  
   
In this paper, we present the first detailed spectral analysis for WR62a. We
demonstrate that it is a short-period SB2. We also report a phase shift in the
orbital solution of one of its emission lines and discuss the
photometric variability of the star.

%__________________________________________________________________

\section{Observations and data reduction}

%_________________________________________________ One column table
\begin{table*}[!t]
\caption{Details of the observations.}
\label{table:1} 
\centering 

\begin{tabular}{c c c c c c c c c c}  
\hline\hline\\ 
     Date-Obs.      & \textit{n} &Sp. coverage&Observat.& Telesc.&Spectr.$^{ a}$& Grating  &Detector& Dispersion & Resolution$^b$\\
   UT         &   & [\AA ]     &         &  [m]   &       &[l~mm$^{-1}$]   &        &[\AA ~pix$^{-1}$]&  \\
\hline
\\
 2002, Apr. 22            & 1 & 3940-5600 & CASLEO & 2.15 & REOSC & 600 & Tek1024      & 1.63& 1000\\
 2007, Mar.  29 - Apr. 3  & 6 & 3650-6700 & CTIO   & 4    & R-C   & 632 & Loral 3k     & 1.01& 1300\\
 2008, Apr. 19-23         & 6 & 3650-6700 & CTIO   & 4    & R-C   & 632 & Loral 3k     & 1.01& 1300\\
 2009, Mar. 24-27         & 4 & 4030-5590 & CASLEO & 2.15 & REOSC & 600 & Tek1024      & 1.63& 1000\\
 2009, Apr. 20-21         & 2 & 3800-5530 & LCO    & 2.5  & B-C   & 1200& Marconi$\#1$ & 0.79& 2500\\
 2009, Jul. 19-27         & 4 & 3520-5250 & LCO    & 2.5  & B-C   & 1200& Marconi$\#1$ & 0.79& 2500\\
 2009, Aug.  14-16        & 3 & 4040-5700 & CASLEO & 2.15 & REOSC & 600 & Tek1024      & 1.63& 1000\\
 2010, Apr.  10-13        & 2 & 3930-5600 & CASLEO & 2.15 & REOSC & 600 & Tek1024      & 1.63& 1000\\
\hline   \\ 
\multicolumn{10}{l}{${n:}$ Number of spectra obtained per run.}\\
\multicolumn{10}{l}{${a:}$ Details of the spectrographs can be found in the User's Manuals of the respective Observatories.}\\
\multicolumn{10}{l}{${b:}$ The spectral resolutions (R=$\lambda / \Delta\lambda$) were measured 
using $\Delta\lambda$ as the FWHM of the calibration lamp emission lines.}\\
\end{tabular}
\end{table*}
%
%______________________________________________________________

We have obtained a total of 28 spectral images of WR62a. 
The images were acquired with the 4-m V. Blanco telescope at Cerro Tololo
  Inter-American Observatory (CTIO), Chile; the 2.5-m du Pont telescope at
Las Campanas Observatory (LCO), Chile; and the 2.15-m J. Sahade telescope at
Complejo Astron\'omico El Leoncito (CASLEO)\footnote{Operated under agreement
  between CONICET and the Universities of La Plata, C\'ordoba, and San
  Juan, Argentina}, Argentina, between 2002 and 2010.
See Table~\ref{table:1} for a summary of the observations. 
We have acquired spectra at CTIO using the Ritchey-Chretien (R-C)
spectrograph, at CASLEO employing the REOSC spectrograph, and at LCO
  using a Boller \& Chivens (B-C) spectrograph. The procedure for obtaining LCO
  spectra are also described in detail by \citet{2011ApJS..193...24S}.   
Slight differences in the spectral ranges listed in Table~\ref{table:1}   
are due to changes in the grating angle among the observing runs.

Typical exposure times for spectra were between 15 and 40 min, resulting in
spectra with signal-to-noise ratios S/N $\sim~$50 -- 150.
Comparison lamp spectra of He-Ne-Ar (or Cu-Ar with REOSC spectrograph) were
observed immediately after or before, at the same telescope position as
the stellar exposures. 
Bias and flat-field images were also obtained each night.
We did not observe a telluric standard star since there are not many
atmospheric absorption lines in the wavelength range of our spectra.
All spectra were processed with IRAF\footnote{IRAF is distributed by the
  National Optical Astronomy Observatories, which are operated by the
  Association of Universities for Research in Astronomy, Inc., under
  cooperative agreement with the National Science Foundation.} routines.

\section{Results and discussion}

\subsection{The spectrum of WR~62a}

Our spectra reveal that WR62a is a double-lined WR+OB binary system. 
Several absorption lines are detected in our spectra. 
The spectra obtained at different epochs show that the absorption lines have different 
radial-velocity shifts, from those of the emission lines, suggesting that the 
system is an SB2.
We used an iterative procedure similar to the one developed by \citet{marchenko_1998}
to separate the composite spectrum into the spectra from the two components of the binary.
First, we begin by shifting the spectra to its respective radial velocity
and calculating the mean spectrum to be used as a template of the WN. 
Second, the RV-shifted mean WR spectrum is subtracted from the original
spectra, obtaining the remaining O-type spectra. 
Then, they are shifted to their corresponding RV of the absorption lines and
combined to get the template of the O-type star.  
After that, the (RV-shifted) mean O-type spectrum is subtracted from the
observed spectra and a new purer WN-type spectrum is obtained. 
That is to say, the spectra are shifted, combined, and subtracted appropriately
to compute the individual spectra of the stellar components. 
We noted that the method converges after a few iterations.

\begin{figure*}[!t]
      \centering
   \includegraphics[width=18cm]{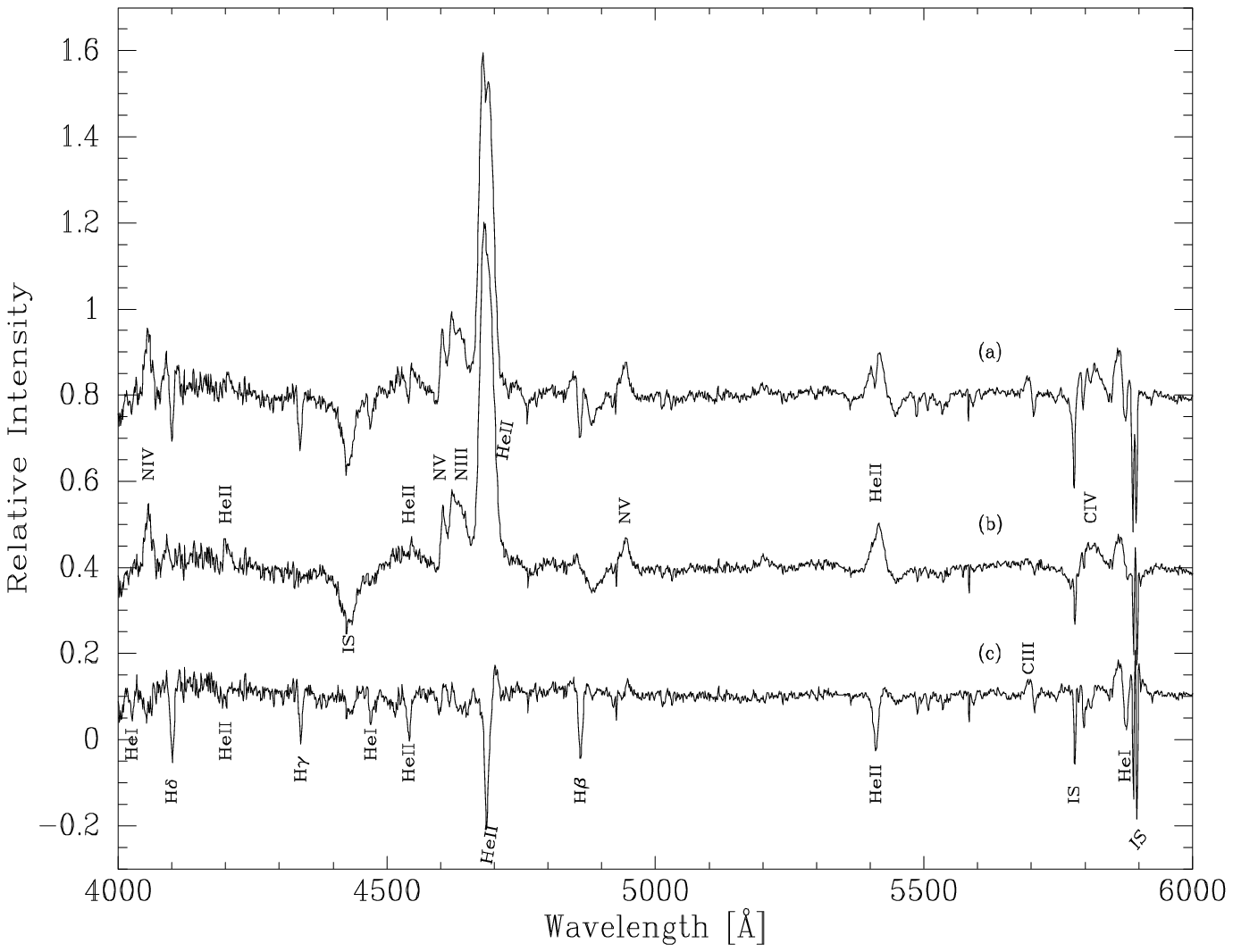}
   \caption{(a) Normalized spectrum of WR62a before the ``disentangling''
     method,  (b) the WR star component after removing the O-type star, (c)
     the remaining spectrum of the O-type star. The spectra are shifted in the
     intensity axis for a better comparison.} 
\label{disentangling}   
       \end{figure*}

After separating the respective spectra (see Fig.~\ref{disentangling}),
we performed a more detailed spectral classification.  
Relative intensities of N\,{\sc iii}, N\,{\sc iv}, He\,{\sc ii}, and C\,{\sc
  iv} emission lines in the WR spectrum indicate a WN5 spectral type
\citep[according to the criteria given by][]{smi96}, confirming the 
classification proposed by \citet{Shara_1999}. 

In the disentangled spectrum of the O-type star, we identified absorption
lines of H$\alpha$, H$\beta$, H$\gamma$, H$\delta$, He\,{\sc i}
$\lambda\lambda$ 4026, 4471, and 5875, and also He\,{\sc ii}
$\lambda\lambda$ 4200, 4542, 4686, and 5411, as well as the C\,{\sc iii} 5696
emission line. 
The comparison between the He\,{\sc i} $\lambda$4471 and the He\,{\sc ii} 
$\lambda$4542 absorption lines indicates that the O-type star is
earlier than O7 spectral type; i.e., the intensity of the former He\,{\sc i}
line is fainter than that of the He\,{\sc ii} line.   
An improved spectral classification was achieved through the comparison 
of the O star spectrum with the MK standards from the new Atlas for spectral
classification \citep{2011ApJS..193...24S}. 
Thus, an O\,5.5-6 type is determined.
A luminosity class is not possible to determine owing to the profile
variations in the emission lines of the WN-type component which introduce
residuals in the resulting O-type spectrum (for example around $4600-4700$\AA).

\subsection{Analysis of radial velocities}

We have determined the RVs of the WN component of WR62a by measuring the
position of the stronger emission lines, i.e. N\,{\sc iv} $\lambda$4058,
N\,{\sc v} $\lambda$4604, and He\,{\sc ii} $\lambda\lambda$4200, 4542,
4686, and 5411. 
For the O-type star, we determined a mean RV value for each 
spectrum by averaging the RV measurements derived from the absorption lines. 
These measurements are listed in Table~\ref{table:2}.  

\begin{figure}[!t]
  \centering
  \includegraphics[width=9cm]{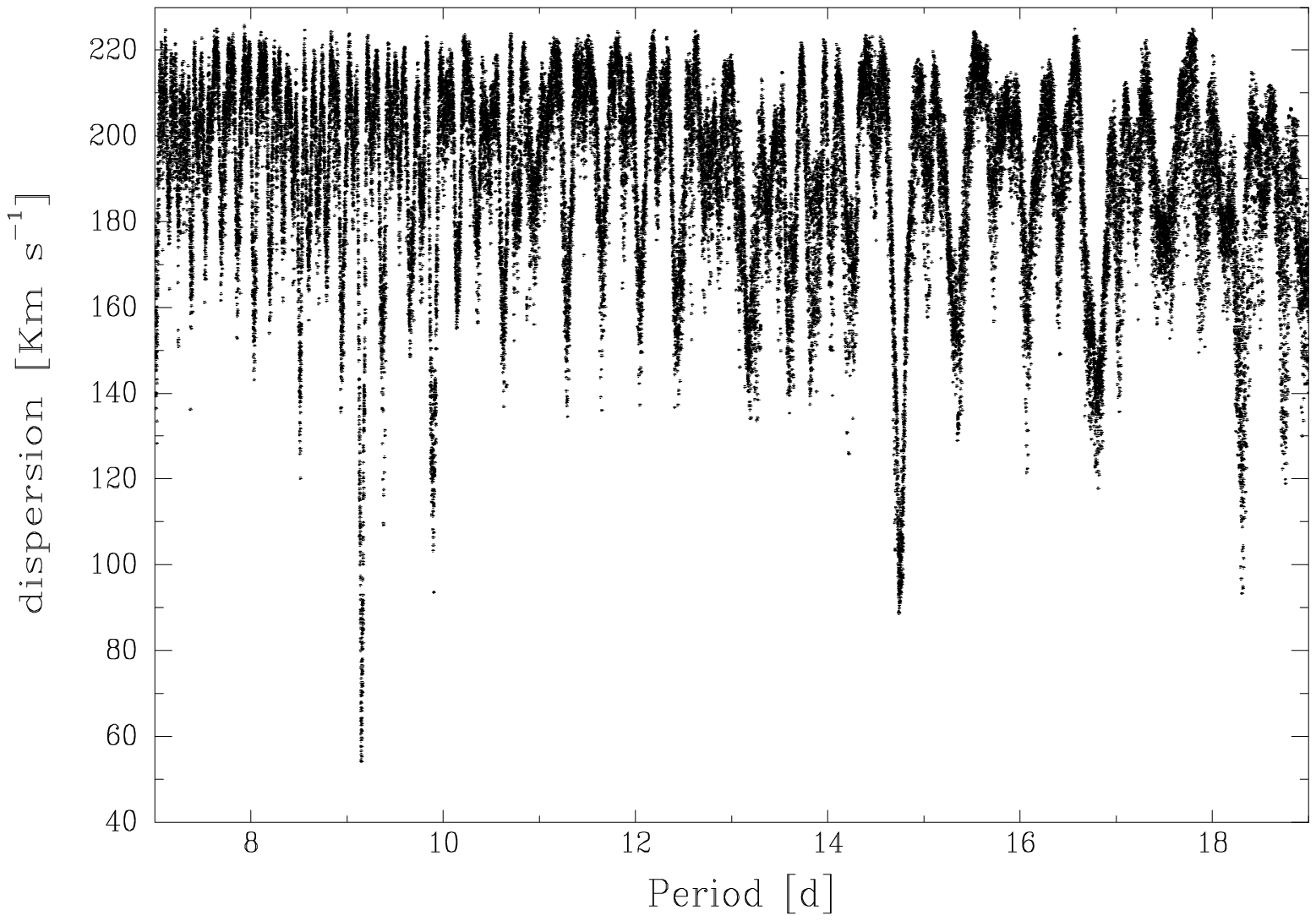}
  \caption{Periodogram obtained for the He\,{\sc ii} $\lambda$4686 emission line
    with the \citet{mar80} code. The dispersion (Y-axis) is the parameter
      used to estimate the quality of the fitting for each trial period. This
      value is the total variance of the best-fitting straight line to the
      observations in each phase interval.}  
  \label{periodograma}
\end{figure}

The strongest emission line in our spectra corresponds to He\,{\sc
  ii} $\lambda$4686, so we could measure this line even in 
spectra with poor signal-to-noise ratios. 
The RVs of this line show large variations from night to night, 
suggesting an orbital period of a few days. 
We have searched for periodicities by means of the algorithm published 
by \citet{mar80}. This method divides the observations into several phase intervals
and computes the variance with respect to the best-fitting straight line in each interval.
 The most probable period obtained corresponds to 9.144 d
(Fig.~\ref{periodograma}), and it is used as ian initial value to feed orbital
solutions for the RV data.

\begin{table*}[!t]
\caption{Observed heliocentric radial velocities of WR62a. 
The mean FWHM and mean EW of each emission line measured from the
  separated spectra are indicated in the last two rows, respectively.}
\label{table:2}  
\centering 
\tiny
\begin{tabular}{c c c c c c c c c c c}  
\hline\hline    \noalign{\smallskip}  
HJD  & Phase    & N\,{\sc iv}  & N\,{\sc v}   & N\,{\sc v}   & C\,{\sc iv}  & 
He\,{\sc ii}    & He\,{\sc ii} & He\,{\sc ii} & He\,{\sc ii} & Mean Abs [n]$^a$\\  
2,450,000+      &  & 4057.66\,\AA & 4603.73\,\AA & 4944.56\,\AA & 5811.98\,\AA &
  4685.68\,\AA  & 5411.52\,\AA & 4199.87\,\AA & 4541.59\,\AA &                 \\
            &   & [km s$^{-1}$]& [km s$^{-1}$]& [km s$^{-1}$]& [km s$^{-1}$]  & 
 [km s$^{-1}$]  & [km s$^{-1}$]& [km s$^{-1}$]& [km s$^{-1}$]& [km s$^{-1}$]  \\

\hline   \noalign{\smallskip}    

2386.72	& 0.98 & -213 &   49 &      &      &  113 &   55 &      &      &  $-43\pm27$ (5) \\
4188.89	& 0.05 &  -65 &      &  -41 &   59 &  254 &  235 &      &      & $-113\pm22$ (8) \\
4189.82	& 0.15 &   53 &  220 &  106 &  155 &  278 &  306 &  308 &  387 & $-204\pm32$ (10)\\
4191.84	& 0.37 &   -8 &  213 &   96 &  185 &  134 &  223 &  297 &  397 & $-175\pm29$ (10)\\
4192.85	& 0.48 & -161 &    9 &   -8 &  -16 &  -61 &   86 &  162 &  261 &  $-83\pm22$ (9) \\
4193.65	& 0.57 & -275 & -119 & -199 & -155 & -162 &  -62 &   42 &  -30 &   $19\pm59$ (9) \\
4193.87	& 0.59 & -228 & -143 & -280 & -135 & -250 & -192 &  -72 &      &   $18\pm52$ (9) \\
4575.84	& 0.36 &   87 &  187 &  102 &  182 &  133 &  259 &  374 &  359 & $-157\pm36$ (10)\\
4576.82	& 0.47 & -122 &   48 &  -32 &  -19 &  -53 &   79 &   90 &      &  $-85\pm28$ (6) \\
4577.73	& 0.57 & -285 & -116 & -193 & -178 & -199 & -119 &      &  -14 &   $19\pm38$ (9) \\
4578.80	& 0.69 & -403 & -254 & -346 & -257 & -282 & -284 &      &      &                 \\
4579.59	& 0.77 & -379 &	-255 & -276 & -257 & -191 & -234 & -246 & -215 &   $31\pm37$ (9) \\
4579.85	& 0.80 &      & -257 &      & -290 & -173 & -250 & -262 & -230 &   $36\pm35$ (9) \\
4914.83	& 0.43 &   -8 &  173 &  101 &      &   49 &  130 &      &  307 &  $-75\pm26$ (5) \\
4915.82	& 0.54 & -134 &  -14 & -161 &      &  -89 &  -25 &  114 &  139 &   $25\pm18$ (3) \\
4916.78	& 0.65 & -255 &      &      &      & -251 & -208 & -181 &  -86 &   $98\pm14$ (4) \\
4917.74	& 0.75 & -406 & -216 & -343 &      & -183 & -282 &      & -212 &   $91\pm23$ (5) \\
4941.57	& 0.36 &   91 &  254 &  150 &      &  218 &  278 &  375 &  362 & $-151\pm24$ (7) \\
4942.71	& 0.48 & -163 &   19 &  -97 &      &  -12 &  -21 &  166 &      &  $-77\pm28$ (8) \\
5031.64	& 0.21 &   81 &  273 &  155 &      &  301 &  302 &  326 &  429 & $-204\pm21$ (8) \\
5036.59	& 0.75 & -372 &      & -288 & -212 & -168 & -227 & -249 & -205 &  $102\pm24$ (7) \\
5038.58	& 0.97 & -204 &  -41 & -167 & -185 &  131 &   33 &  -32 &  109 &  $-97\pm36$ (3) \\
5039.61	& 0.08 &  -31 &  123 &   -1 &      &  263 &  218 &  228 &  244 & $-168\pm33$ (7) \\
5057.52	& 0.04 &      &  170 & -118 &      &  352 &  314 &      &      & $-137\pm40$ (7) \\
5059.52	& 0.25 &      &  287 &  175 &      &  356 &      &      &      & $-183\pm28$ (6) \\
5060.49	& 0.36 &      &  191 &   60 &      &  187 &  237 &      &      & $-151\pm17$ (7) \\
5296.79	& 0.20 &  153 &      &      &      &  267 &      &      &  373 & $-195\pm33$ (6) \\
5299.66	& 0.51 &      &      &      &      & -113 &      &      &      &   $31\pm1$  (4) \\

\hline   \noalign{\smallskip} 

{\em FWHM} [\AA]             & &  18.5 & 10.4 & 19.7 & 33.9 &  24.7 & 22.4 & 17.3 & 18.7 & \\
{\em EW} [\AA]               & &  -4.6 & -1.6 & -1.5 & -2.2 & -26.1 & -3.8 & -1.2 & -1.6 &  \\

\hline                    
\hline   \noalign{\smallskip}  
\multicolumn{11}{l}{$a$: 
the errors are the standard deviation of the mean, 
and the (n) values indicate the number of absorption 
lines measured.} 
\end{tabular}
\end{table*}
  
\begin{figure}[!t]
  \centering
  \includegraphics[width=9cm]{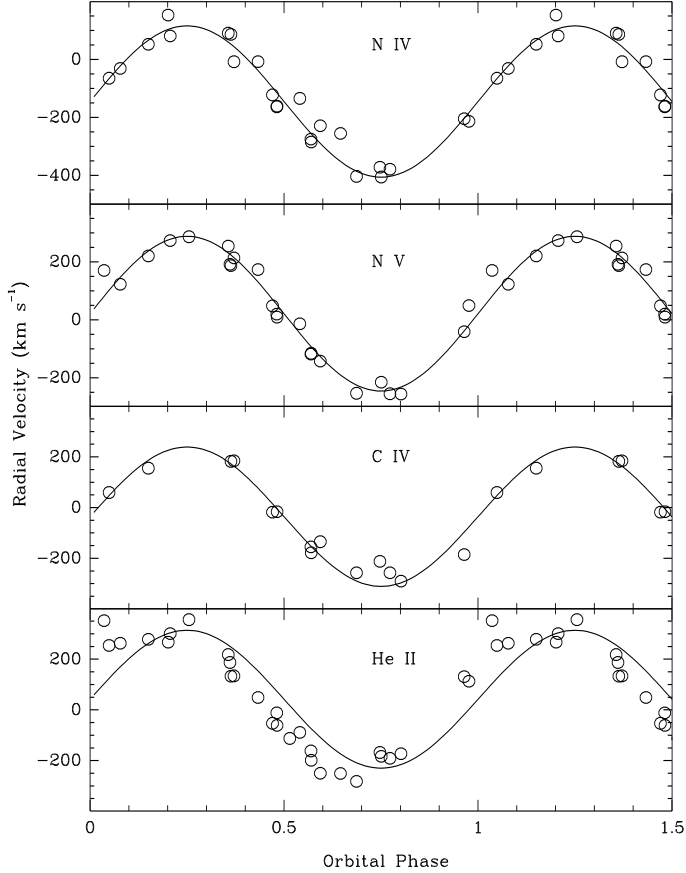}
  \caption{
Radial velocities of N\,{\sc iv} $\lambda$4058 , N\,{\sc v}~$\lambda$4604, C\,{\sc iv}
$\lambda$5812, and He\,{\sc ii} $\lambda$4686
emission lines phased with the ephemeris of Eq. 1. Continuous curves represent
the orbital solutions from Table~\ref{table:3}.  The phase shift of the
  He\,{\sc ii} line is noticeable.}  
  \label{vrs}
\end{figure}

We performed orbital fits to each dataset of RVs, using the 
{\sc gbart}\footnote{{\sc gbart} is an improved version of the program
  originally written by \citet{ber68} and developed by Federico Bareilles.} program.
It is clear that the determined RVs for the different spectral features show
the same periodicity (see Table~\ref{table:P}), and thus, 
we adopted the straight mean of the periods as the definitive value for the orbital 
period. The ephemeris adopted for WR62a is

\begin{equation}
T_{\rm 0}=2,455,038.9+9.1447 d \times E 
\end{equation}

\noindent
where $T_{\rm 0}$ is the time when the WN star passes in front of the system,
as in the normal convention for eclipsing WR+OB systems.
Then, we calculated the orbital solutions with the period fixed to this average.
The orbital parameters derived for each RV dataset are presented in
Table~\ref{table:3} (with their corresponding probable errors).
To determine the $T_{\rm 0}$ of the system, we neglected the very low
eccentricity, averaging all the $T_{\rm RVmax}$ values of the emission lines of
Table~\ref{table:3} and calculating the 
$T_{\rm 0}$ as $ < T_{\rm RVmax} > + 0.75 \times P$, 
which coincides with the primary minimum of the lightcurve 
(as will be shown in Sect. \ref{foto}). Some orbital solutions are
illustrated in Fig.~\ref{vrs}. The RV curves are phased with the ephemeris
from Eq.~1. 

In Fig.~\ref{vrs} and Table~\ref{table:3}, a phase shift between the
He\,{\sc ii} $\lambda$4686 line and other emission lines is noticeable. 
As a result, the $T_{\rm RVmax}$ value for this line was excluded from the calculus
of the ephemeris.
This phase-shift effect is also detected in other close WR binary systems, 
e.g. 
HD~90657   \citep[\object{WR 21}; ][]{nie82,gam04}, 
HD~94546   \citep[\object{WR 31}; ][]{gam04}, and 
HDE~320102 \citep[\object{WR 97}; ][]{nie95,gam04}, 
and usually interpreted as a sign of colliding winds.
The amplitudes and systemic velocities of the orbital motion determined from the RV of 
different emission lines show differences. 
The amplitude of N\,{\sc v} $\lambda$4944 shows the lowest value, 
while He\,{\sc ii} $\lambda\lambda$4200, 4541, and 5411 emission 
lines appear somewhat larger. 
The systemic velocities of emission lines (except N {\sc iv} $\lambda4058$ and 
N {\sc v} $\lambda4944$) are redshifted relative to that of the absorption lines.
This is a known phenomenon in WR+OB systems, which
indicates that emission lines are being formed in different regions of the WN
wind \citep[cf. ][]{nie82,nie95}, or perhaps are still affected by the absorption lines of the 
secondary.

\begin{table}
  \caption{Periodicities 
      derived from the radial velocity of N\,{\sc iv},  N\,{\sc v},
      C\,{\sc iv}, He\,{\sc ii} emission lines and the mean absorption.}
  \label{table:P}
  \centering
  \tiny
  \begin{tabular}{l r l} 
  \hline\hline    \noalign{\smallskip}
Feature                    & P [d]  &error [d]\\
\hline    \noalign{\smallskip}
N\,{\sc iv} $\lambda$4058  & 9.1444 &$0.0005$\\
N\,{\sc v} $\lambda$4604   & 9.1455 &$0.0005$\\
N\,{\sc v} $\lambda$4944   & 9.1443 &$0.0009$\\
C\,{\sc iv} $\lambda$5812  & 9.146  &$0.002$ \\
He\,{\sc ii} $\lambda$4686 & 9.1447 &$0.0005$\\
He\,{\sc ii} $\lambda$5411 & 9.1447 &$0.0003$\\
He\,{\sc ii} $\lambda$4200 & 9.145  &$0.001$ \\
He\,{\sc ii} $\lambda$4542 & 9.140  &$0.002$ \\
Mean Abs.                  & 9.1415 &$0.0005$\\
\hline 
\end{tabular}
\end{table}

\begin{table*}
  \caption{ The orbital parameters (with their respective probable errors)
      derived from the radial velocity curves of N\,{\sc iv},  N\,{\sc v},
      C\,{\sc iv}, He\,{\sc ii} emission lines, and the mean absorption.}
  \label{table:3}      
  \centering       
  \tiny  
  \begin{tabular}{c c c c c c c c c c} 
  \hline\hline    \noalign{\smallskip}   
Parameter & N\,{\sc iv} & N\,{\sc v} & N\,{\sc v} & C\,{\sc iv} & He\,{\sc ii}  & He\,{\sc ii}  &  He\,{\sc ii}  &  He\,{\sc ii} & Mean Abs.  \\ 
          & 4058        & 4604       & 4944       & 5812        & 4686          & 5411          &  4200          &  4542         &            \\     
\hline \noalign{\smallskip}
P [d]  &  \multicolumn{9}{c}{9.1447$\pm$0.0014} \\                   

V$_0$ [km~s$^{-1}$]        & $-145\pm5$	   & $21\pm5$       & $-94\pm5$      & $-36\pm7$      &	$42\pm4$       & $47\pm4$       & $80\pm6$       & $133\pm6$      & $-65\pm3$      \\
K [km~s$^{-1}$]	          & $261\pm7$      & $267\pm7$      & $246\pm7$      & $275\pm13$     & $272\pm6$      & $303\pm5$      & $321\pm8$      & $320\pm8$      & $143\pm5$	   \\
\textit{e}                & $0.07\pm0.02$  & $0.05\pm0.03$  & $0.08\pm0.02$  & $0.11\pm0.04$  & $0.07\pm0.02$  & $0.09\pm0.02$  & $0.08\pm0.02$  & $0.11\pm0.03$  & $0.08\pm0.04$  \\
$\omega$ [deg]            & $12\pm22$      & $160\pm29$     & $24\pm20$      & $-14\pm17$     & $159\pm16$     & $183\pm11$     & $166\pm19$     & $160\pm14$	  & $319\pm21$     \\
T$_{\rm RVmax}$ [d]$^{\ast}$  & $5032.1\pm0.5$ & $5032.0\pm0.7$ & $5032.2\pm0.5$ & $5032.1\pm0.4$ & $5031.4\pm0.4$ & $5031.8\pm0.3$ & $5032.3\pm0.5$ & $5032.2\pm0.4$ & $5036.1\pm0.5$ \\

 \hline  \noalign{\smallskip}
 \multicolumn{10}{l}{$\ast$ T$_{\rm RVmax}$ = Time of maximum RV in Heliocentric Julian days $-2,450,000$}\\
 \end{tabular}
\end{table*}

\begin{figure}[!t]
  \centering
  \includegraphics[width=9cm]{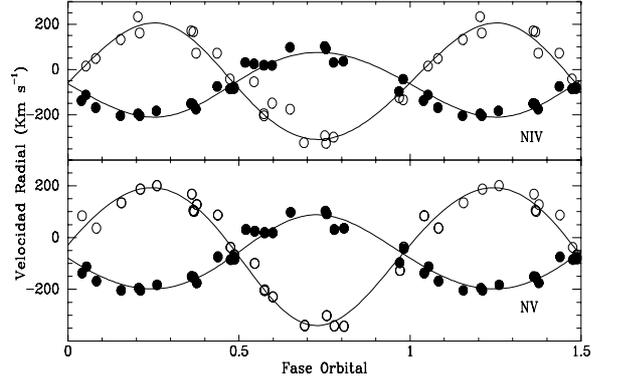}
  \caption{ The RV curves for both components in WR62a. 
  The open circles show the RVs for N\,{\sc iv}~$\lambda$4058 \AA\ and 
  N\,{\sc v}~$\lambda$ 4604 \AA\ emission lines.  
  The filled circles show the mean RVs of some absorption lines for the
  secondary one.  
  RVs of the emission lines were shifted to match the systemic velocity
  derived from the absorption lines alone (see text). 
  The ephemeris given in Eq.1 was used to generate the RV curves.}   
  \label{fig:sb2}
\end{figure}

From Table~\ref{table:3} and Fig.~\ref{fig:sb2} it is also evident that the
absorption lines (the mean of H, He\,{\sc i}, He\,{\sc ii}) move in anti-phase
with the emissions, thus they belong to a secondary component.   
Besides, the semi-amplitude of the orbit of this companion is lower than the
WN one (regardless of emission line used), which means that the secondary is more
massive than the WN component. 
 
We performed fits of the orbital solution for both components together, by
means the {\sc gbart} program. 
It is widely known that the RVs of the different emission lines do not strictly follow
the same orbital motion in WR binary systems, which is often
attributed to their different line-forming regions.  
We assumed that the orbital motion of the WN star is better represented
by the RVs of the highest ionization emission lines, namely, 
N\,{\sc iv}~$\lambda$ 4058 \AA\ and N\,{\sc v}~$\lambda$ 4604 \AA. 
The O-type star is depicted by the RVs of the mean of some absorption lines.
To perform the simultaneous fit of both RV curves, we shifted
the emission-line RVs to match the systemic velocity derived for the O-type
component.  
Therefore, the RVs of N\,{\sc iv} and N\,{\sc v} were shifted by
80~km\,s$^{-1}$ and $-86$~km\,s$^{-1}$, respectively 
(see Table~\ref{table:3}).  
These solutions are shown in Table~\ref{table:4} and Fig.~\ref{fig:sb2}.
Thus, the secondary component is confirmed as the more massive star.
Depending on which orbital solution is considered, the mass of the O-type star ranges
between 39 and 42 $M_\odot$, and the WN mass, between 21 and 23 $M_\odot$. The
mass ratio ($q$) obtained agrees with the values determined in other WN5+OB
systems \citep{huc01}.   
The low eccentricity determined using different datasets indicates that the
orbit is almost circular, although such a low value should not be neglected.

\begin{table}
\tiny
\caption{Orbital solutions for both component of WR~62a}             
\label{table:4}      
\centering     
     
\begin{tabular}{c c c } 
\hline\hline   \noalign{\smallskip}  
                         & N\,{\sc iv}+abs. &  N\,{\sc v}+abs.	\\
\hline   \noalign{\smallskip}

$P$ [d]                   &  \multicolumn{2}{c}{9.1447 (fixed)} \\ 
$e$                       & $0.05\pm0.02$ & $0.05\pm0.02$ \\
$K_{\rm WR}$ [km~s$^{-1}$]   & $257\pm7$  &    $266\pm6$	\\
$K_{\rm O}$ [km~s$^{-1}$]	  & $143\pm6$  & $143\pm6$	\\
$V_0$ [km~s$^{-1}$]        & $-62\pm3$  & $-62\pm3$      \\
$\omega$ [deg]            & $39\pm25$ & $149\pm20$ \\
$T_{\rm Periast}$ [d]$^{*}$       & $5,032.9\pm0.6$ & $5,035.7\pm0.5$ \\
$T_{\rm RVmax}$ [d]$^{*}$         & $5,032.0\pm0.6$ & $5,031.9\pm0.5$ \\
$a_{\rm WR}\sin i$ [$R_\odot$]   & $46.1\pm1.1$   	  &    $47.8\pm1.1$   \\
$a_{\rm O} \sin i$ [$R_\odot$]   & $25.6\pm1.1$	  &    $25.7\pm1.1$    \\
$M_{\rm WR}\sin^{3}i$ [$M_\odot$] & $21.5\pm4.8$ & $22.6\pm5.0$		\\
$M_{\rm O}\sin^{3}i$  [$M_\odot$] & $38.7\pm5.2$ & $42.0\pm5.0$	\\
\textit{q({\rm WR}/{\rm O}}   &  $0.56\pm0.04$  & $0.54\pm0.04$ \\
\hline  \noalign{\smallskip}
\multicolumn{3}{l}{$\ast$ HJD-2,450,000}\\
\end{tabular}
\end{table}

\subsection{Analysis of spectral features}

\begin{figure}[!t]
      \centering
      \includegraphics[width=9.5cm,height=5cm]{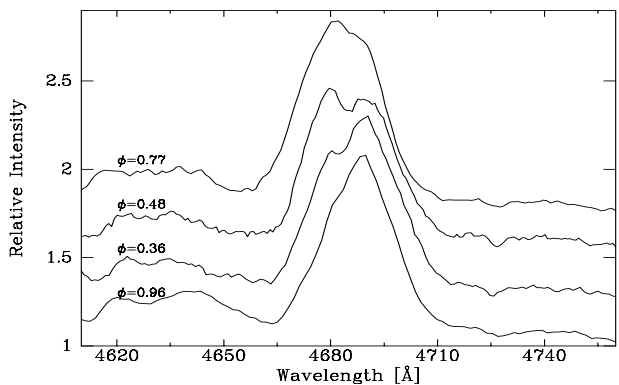}
      \caption{Continuum rectified spectra of the He\,{\sc ii} $\lambda$4686 emission
        line with the superimposed absorption observed during different
        orbital phases of the binary. 
        A noticeable anti-phased movement of absorption and emission features is shown.}   
      \label{he2}
\end{figure}

\begin{figure}[!t]
    \centering
    \includegraphics[width=9cm]{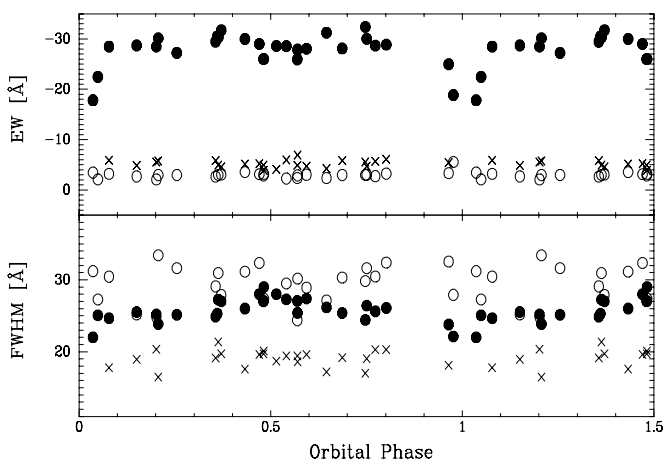}
    \caption{{\em EW} and {\em FWHM} measured in the He\,{\sc ii} $\lambda$4686 (filled circles), 
    N\,{\sc iv} $\lambda$4058 (crosses), and N\,{\sc v} $\lambda$4944 (open circles) emission lines.}    
    \label{ew}
\end{figure}

In Fig.~\ref{he2}, we show the He\,{\sc ii} $\lambda$4686, line at four
different orbital phases; the absorption line of the secondary can be seen
to be moving in anti-phase with respect to the emission line peak.
In some phases, this emission line shows an asymmetric profile, perhaps
indicating some additional contribution from the wind-wind colliding region.
We measured the equivalent width ({\em EW}) and full-width at half-maximum 
({\em FWHM}) of the He\,{\sc ii} $\lambda$4686, N\,{\sc iv} $\lambda$4058,
and N\,{\sc v} $\lambda$4944 emission lines. 
We detected a certain variability of these line parameters, which seems to be modulated through
the orbit (see Fig.~\ref{ew}). 
The behaviour of the {\em EW} of the He\,{\sc ii} line is very noticeable: 
it decreases around phase $\phi$=0.0, when the WN star is in front of the system.  
This phenomenon is also found in other WR+OB binary systems, i.e. WR21 and
WR47 \citep{gam04}, and HD~5980 \citep{1998ApJ...497..896M},
explained as if the emission line had an additional non-stellar component, 
originating in the colliding-winds region, which is eclipsed during the conjunction phase.

\subsection{Analysis of the available photometry}
\label{foto}

\begin{figure}[!t]
   \centering
   \includegraphics[width=9cm]{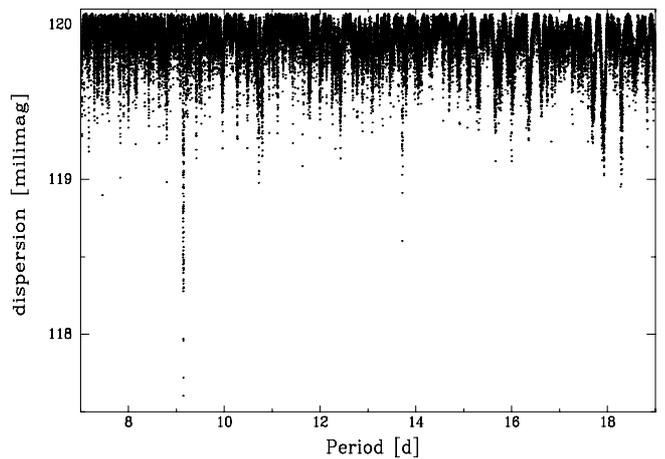}
   \caption{Periodogram obtained from the ASAS data with the \citet{mar80} code. 
   The most probable period found is 9.1438$\pm$0.0009 d.}   
   \label{marmuzASAS}
\end{figure}

\begin{figure}[!t]
  \centering
  \includegraphics[width=9cm]{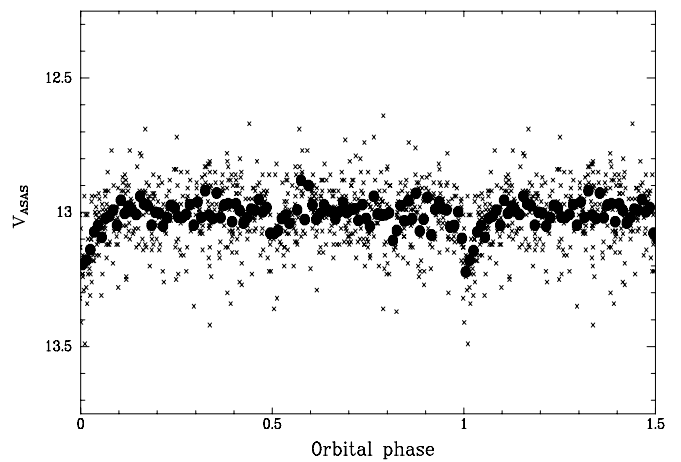}
  \caption{The lightcurve of WR62a using the ASAS data. 
The individual data-points are shown by crosses; the filled circles represent
the data-points averaged in phase bins of 0.01. The ephemeris derived 
from our spectroscopic data (Eq. 1) is used to calculate the orbital phases.}
  \label{asas}
\end{figure}

As the SB2 orbital solution for WR62a suggests high minimum masses for both
components of the system and the {\em EW} of the He\,{\sc ii} $\lambda$4686
emission line has a minimum at one of the conjunctions, we would expect to
observe photometric variability (eclipses, for instance). 
We, therefore, analysed the data published in the ``All-Sky Automated
Survey'' \citep[ASAS; ][]{pojmanski_2001}, which gives the {\em V} magnitudes
for this star.   
Again, using the code of \citet{mar80}, we searched the ASAS data for
periodicities and obtained the most probable period of 9.1438$\pm$0.0009 d,
very similar to the period obtained from our RV analysis (see the
periodogram in Fig.~\ref{marmuzASAS}). In Fig.~\ref{asas}, we have plotted
the ASAS $V$ magnitudes against the orbital phases generated using the
ephemeris given in Eq. 1.  
We see a V-shaped dip at $\phi=0.0$. A much shallower dip is marginally
detected at $\phi=0.5$. 
To improve the visualization of the variations, we also show
the data points averaged in phase bins of 0.01 in Fig.~\ref{asas}. 
The shape of the dip at $\phi=0.0$ calls the non-photospheric eclipsing
binaries analysed by \citet{lam96}. In that paper, the authors analysed
the eclipsing binary systems WR21 and WR47, where, as we described in
the previous sections, the He\,{\sc ii} $\lambda$4686 emission line also shows 
a phase shift in RVs and a minimum of the {\em EW} when the WN star passes in
the front of the system.
The scenario where an additional emission component to the He\,{\sc ii}
$\lambda$4686 emission is formed in a colliding-wind region, which is
being eclipsed by the WN component, should also apply.
However, the accuracy of the ASAS data does not allow us to determine the 
inclination of the system, hence the absolute masses for both components.
The {\em V} magnitude of WR62a is close to the faint limit of the ASAS survey. 
More accurate photometry is required to confirm or rule out the eclipsing
nature of this binary.

\section{Conclusions}

From our spectroscopic observations and analysis of the RV of the lines
detected in the spectra of WR62a, we discovered that it is a double-lined
binary system with a 9.1447 d orbital period.  
The relative intensities of the emission lines detected in our spectra confirm
that the spectral classification of the WN star as WN5, while the absorption lines
identified for the secondary star corresponds to an O\,5.5-6 spectral-type. 
The emission lines of N, C, and He ions originated in the WR star follow the
same movement and orbital period, but we detect a slight phase shift in the 
He\,{\sc ii} $\lambda$4686 emission line, which may be due to
asymmetries in the line-forming region.  

We performed fits of the orbital solution for both components by considering two
possible RV datasets of the WN  star. 
From these solutions, we confirm that the O-type star is more massive than the
WN star. 
The minimum mass derived for the O-type  star ranges from 39 to 42
${M}_\odot$, and for the WN-type star from 21 to 23~${M}_\odot$, 
depending on which emission line is used to represent the orbital motion of the WN
star, i.e. N\,{\sc iv} $\lambda$4058 or N\,{\sc v} $\lambda$ 4604 respectively. 

We detected variations in the {\em EW} of the He\,{\sc ii} $\lambda$4686
emission line with the orbital period, showing a $\sim$30\% ``dip''
around $\phi=0$, where the WN star is in front.
This effect suggests that an additional emission component to this line,
originated in a colliding-wind zone, is being eclipsed. 
Analysing photometric data from ASAS, we also identified a minimum when
the WN star is in front of the system. Additionally, a very marginal secondary 
dip is detected in the opposite orbital phase (around $\phi=0.5$). 
The quality of ASAS data does not allow us to determine the inclination of the 
system hence the absolute masses for both components. 
To confirm the eclipsing nature of WR62a, we need more accurate photometry. 

The spectrum of WR62a shows a remarkable resemblance to
the WR+O binary system WR21 \citep[]{gam04}. 
Similar phenomena are found in both binary systems. 
These systems could be used as test benches for geometric colliding-wind models.

\begin{acknowledgements}
We are very grateful to the referee for the constructive comments
We thank the directors and staff of CASLEO, LCO, and CTIO for allowing us
the use of their facilities. 
We especially thank Nidia ``Hada'' Morrell for kindly obtaining some spectra.
%RHB acknowledges support from FONDECYT Project No. 1120668.
\end{acknowledgements}

{\em Facilities:} {CTIO: 4m, LCO: 2.5m, CASLEO: 2.15m.}

\bibliographystyle{aa}
\bibliography{aa_wr62a}

\begin{thebibliography}{18}
\expandafter\ifx\csname natexlab\endcsname\relax\def\natexlab#1{#1}\fi

\bibitem[{{Barb{\'a}} {et~al.}(2010){Barb{\'a}}, {Gamen}, {Arias}, {Morrell},
  {Ma{\'{\i}}z Apell{\'a}niz}, {Alfaro}, {Walborn}, \&
  {Sota}}]{2010RMxAC..38...30B}
{Barb{\'a}}, R.~H., {Gamen}, R., {Arias}, J.~I., {et~al.} 2010, in Revista
  Mexicana de Astronomia y Astrofisica Conference Series, Vol.~38, Revista
  Mexicana de Astronomia y Astrofisica Conference Series, 30--32

\bibitem[{Bertiau \& Grobben(1968)}]{ber68}
Bertiau, F. \& Grobben, J. 1968, Ric. Astr. Spec. Vat., 8, 1

\bibitem[{{Gamen}(2004)}]{gam04}
{Gamen}, R. 2004, Ph.D. Thesis, La Plata University

\bibitem[{{Lamontagne} {et~al.}(1996){Lamontagne}, {Moffat}, {Drissen},
  {Robert}, \& {Matthews}}]{lam96}
{Lamontagne}, R., {Moffat}, A.~F.~J., {Drissen}, L., {Robert}, C., \&
  {Matthews}, J.~M. 1996, \aj, 112, 2227

\bibitem[{{Langer} \& {Heger}(1999)}]{Langer_1999}
{Langer}, N. \& {Heger}, A. 1999, in IAU Symposium, Vol. 193, Wolf-Rayet
  Phenomena in Massive Stars and Starburst Galaxies, ed. K.~A. {van der Hucht},
  G.~{Koenigsberger}, \& P.~R.~J. {Eenens}, 187--+

\bibitem[{{Marchenko} {et~al.}(1998){Marchenko}, {Moffat}, \&
  {Eenens}}]{marchenko_1998}
{Marchenko}, S.~V., {Moffat}, A.~F.~J., \& {Eenens}, P.~R.~J. 1998, \pasp, 110,
  1416

\bibitem[{Marraco \& Muzzio(1980)}]{mar80}
Marraco, H. \& Muzzio, J. 1980, PASP, 92, 700

\bibitem[{{Meynet} \& {Maeder}(2005)}]{Meynet_2005}
{Meynet}, G. \& {Maeder}, A. 2005, \aap, 429, 581

\bibitem[{{Moffat} {et~al.}(1998){Moffat}, {Marchenko}, {Bartzakos}, {Niemela},
  {Cerruti}, {Magalhaes}, {Balona}, {St-Louis}, {Seggewiss}, \&
  {Lamontagne}}]{1998ApJ...497..896M}
{Moffat}, A.~F.~J., {Marchenko}, S.~V., {Bartzakos}, P., {et~al.} 1998, ApJ,
  497, 896

\bibitem[{{Niemela} {et~al.}(1995){Niemela}, {Cabanne}, \& {Bassino}}]{nie95}
{Niemela}, V.~S., {Cabanne}, M.~L., \& {Bassino}, L.~P. 1995, Revista Mexicana
  de Astronom\'{\i}a y Astrof\'{\i}sica, 31, 45

\bibitem[{Niemela \& Moffat(1982)}]{nie82}
Niemela, V.~S. \& Moffat, A. 1982, Ap.J., 259, 213

\bibitem[{{Pojma{\'n}ski}(2001)}]{pojmanski_2001}
{Pojma{\'n}ski}, G. 2001, in Astronomical Society of the Pacific Conference
  Series, Vol. 246, IAU Colloq. 183: Small Telescope Astronomy on Global
  Scales, ed. {B.~Paczynski, W.-P.~Chen, \& C.~Lemme}, 53--+

\bibitem[{{Sana} \& {Evans}(2011)}]{2011IAUS..272..474S}
{Sana}, H. \& {Evans}, C.~J. 2011, in IAU Symposium, Vol. 272, IAU Symposium,
  ed. {C.~Neiner, G.~Wade, G.~Meynet, \& G.~Peters}, 474--485

\bibitem[{{Shara} {et~al.}(1999){Shara}, {Moffat}, {Smith}, {Niemela},
  {Potter}, \& {Lamontagne}}]{Shara_1999}
{Shara}, M.~M., {Moffat}, A.~F.~J., {Smith}, L.~F., {et~al.} 1999, \aj, 118,
  390

\bibitem[{Smith {et~al.}(1996)Smith, Shara, \& Moffat}]{smi96}
Smith, L., Shara, M., \& Moffat, A. 1996, MNRAS, 281, 163

\bibitem[{{Sota} {et~al.}(2011){Sota}, {Ma{\'{\i}}z Apell{\'a}niz}, {Walborn},
  {Alfaro}, {Barb{\'a}}, {Morrell}, {Gamen}, \& {Arias}}]{2011ApJS..193...24S}
{Sota}, A., {Ma{\'{\i}}z Apell{\'a}niz}, J., {Walborn}, N.~R., {et~al.} 2011,
  \apjs, 193, 24

\bibitem[{{Usov}(1991)}]{1991MNRAS.252...49U}
{Usov}, V.~V. 1991, \mnras, 252, 49

\bibitem[{{van der Hucht}(2001)}]{huc01}
{van der Hucht}, K.~A. 2001, New Astronomy Reviews, 45, 135

\end{thebibliography}

\end{document}